# THE RELATIVISTIC SPECTROSCOPY OF ROTATING BODIES

S.V. Artemov, M.A. Belov, Yu.N. Koblik, G.A. Radyuk

*Institute of Nuclear Physics of Uzbek Academy of Sciences, pos. Ulugbek, Tashkent, 702132, Uzbekistan. E-mail:* **belov@uzsci.net**

Abstract

Mutual dependencies of mass and orbital moment increase for rotating bodies of various configurations (a string with uniform mass distribution, a string with squared mass distribution or a sphere with uniform mass distribution, double system of point masses) are considered in quasiclassical approximation. Results are used for analysis of meson, baryon resonances of rotating nature, and also for dipion and dinucleon resonances. For oneparticle resonances a string configuration with squared mass distribution is has been chosen. Radius values and their possible changes with spin increase are obtained. At this, in particular, for proton and Δ-isobar, which are the founders of rotating bands, values 0.45 and 0.63 are defined, respectively. Dipion and dinucleon resonances are presented as two rotating point bodies at distances of some fermy. Each of resonances is located in four rotary bands. Possible reasons of formation of some rotary bands for these resonances are discussed.

## 1. Introduction

In nuclear and molecular physics and in the theory of elementary particles two basic types of dependence between state energies and moments of rotating system (body) are known. At low energies, when the state energy is not great in comparison with the rest mass M (rotating molecules and atomic nuclei), the state energy E, radius R and the moment is connected by the equation

$$E = M + \frac{c^2 \hbar^2}{2MR^2} \cdot J(J+1). \tag{1}$$

Here M - the lowest one from the series of rotating states.

In the other extreme case, when objects rotate with large speeds (baryon resonances), the linear dependence of the orbital moment on square of the state energy is observed:

$$J = \alpha \cdot E^2 + const. \tag{2}$$

This dependence corresponds to the Chu-Frauchi diagram or (originally) – the Redje trajectory and is widely used at the analysis of baryon supermultiplets. At the analysis of electron-positron and dipion resonances we deal with state energies, extending from the rest mass up to values, in some times exceeding the rest mass. Therefore, for the analysis of such resonances there was a requirement for working out of relativistic spectroscopy of rotating bodies.

## 2. Resonance masses and orbital moments of rotating bodies

### 2.1. The rotating string with the constant density of mass on the length unit

By the analogy with /1/ we will find an expression of energy and angular moment, depending on the speed of rotation. The axis of rotation makes a right angle with a longitudinal axis of a body. Let $k$ be the density of energy on the length unit. The local speed $v$ in a point at distance $r$ is defined from the equation

$$v/c = \beta \cdot r/R,$$

where $R$- a half length of a string, and $\beta$ - a parameter of rotating velocity. The points have maximal velocity at body ends and here $\beta$ is the relativistic factor. For all others body points it jointly with the ratio $r/R$ defines the local velocity of the point.

The total relativistic mass is:

$$E = 2\int_0^R \frac{k \mathrm{d}r}{\sqrt{1-v^2/c^2}} = 2Rk \cdot \frac{\arcsin\beta}{\beta} = M \frac{\arcsin\beta}{\beta}. \tag{3}$$

At the limit of small speeds at $\beta = 0$ the body energy is equal to its initial mass: $E_{min} = M = 2\kappa R$.

In extremely relativistic case at $\beta = 1$ a body has the greatest possible mass, which exceeds the rest mass in $\pi/2$:

$E_{max} = M \cdot \pi/2$.

Let us find now an orbital angular moment:

$$|\vec{J}| = \frac{2}{\hbar c^2} \int_0^R \frac{krv dr}{\sqrt{1 - v^2/c^2}} = \frac{kR^2}{c\hbar}\left(\frac{\arcsin\beta - \beta\sqrt{1-\beta^2}}{\beta^2}\right). \quad (4)$$

For reposing body the orbital moment is equal to zero, and its maximal value at $\beta = 1$ is equal

$$|\vec{J}| = \frac{M \ R}{2c\hbar} \cdot \frac{\pi}{2}.$$

Using (3), (4) and expression of the module of orbital moment for quantum objects we will receive required correlation for a rotational band:

$$\frac{R}{2\hbar c}\left(\frac{E - M\sqrt{1-\beta^2}}{\beta}\right) = \sqrt{J(J+1)}. \quad (5)$$

Expanding expressions (3) and (4) in a series of small $\beta$:

$$E \to M\left(1 + \frac{1}{6}\beta^2 + ...\right)$$

$$|\vec{J}| \to \frac{RM}{\hbar c}\left(\frac{1}{3}\beta + ...\right)$$

and excluding $\beta$, we receive a correlation between $E$ and $J$, similar to (1):

$$E \cong M + \frac{3}{2}\cdot\frac{\hbar^2 c^2}{MR^2}\cdot J(J+1).$$

At large speeds, at $\beta > 0.9$, when $E/M > 1.25$, as calculations show, the following dependence between $E$ and $J$ (see fig. 1) is observed:

$J = \alpha E^2 + const$.

## 2.2. The string with the squared distribution of density

Calculating integrals in (3) and (4) for $\kappa = \mu R^2$ we will receive next expressions for the energy and moment:

$$E = \mu \cdot R^3 \cdot \left(\frac{\arcsin\beta - \beta\cdot\sqrt{1-\beta^2}}{\beta^3}\right) = \frac{3}{2} M \cdot \left(\frac{\arcsin\beta - \beta\cdot\sqrt{1-\beta^2}}{\beta^3}\right) \quad (6)$$

$$|\vec{J}| = \frac{3}{4}\cdot\frac{\mu R^4}{\hbar c}\cdot\left(\frac{\arcsin\beta - \beta\cdot\sqrt{1-\beta^2} - \frac{2}{3}\beta^3\cdot\sqrt{1-\beta^2}}{\beta^4}\right) \quad (7)$$

Their limiting values are equal:

$$E = \begin{cases} \frac{2}{3}\mu \cdot R^3 \equiv M & \text{at } \beta = 0 \\ M \cdot \frac{\pi}{2}\cdot\frac{3}{2} = M \cdot 2{,}355 & \text{at } \beta = 1 \end{cases},$$

$$|\vec{J}| = \begin{cases} 0 & \text{at } \beta = 0 \\ \dfrac{MR}{\hbar c} \cdot \dfrac{\pi}{2} \cdot \dfrac{9}{8} & \text{at } \beta = 1 \end{cases}.$$

Similarly to item 2.1. it is possible to receive a correlation between energy and angular moment, distinguished from (5) by numerical multiplier:

$$\frac{3}{4} \cdot \frac{R}{\hbar c}\left(\frac{E - M\sqrt{1-\beta^2}}{\beta}\right) = \sqrt{J(J+1)}. \qquad (8)$$

At small β, acting again as in item 2.1., we receive:

$$E \to M\left(1 + \frac{3}{10}\beta^2 + \ldots\right),$$

$$|\vec{J}| \to \frac{RM}{\hbar c}\left(\frac{3}{5}\beta + \ldots\right)$$

and the correlation, similar to (1):

$$E \cong M + \frac{5}{6} \cdot \frac{c^2 \hbar^2}{MR^2} \cdot J(J+1).$$

At large velocities, at β > 0.9, when $E/M$ > 1.5, as computations show, the following dependence between $E$ and $J$ is observed
$J = \alpha E^{1.3} + const$.

## 2.3. Two point objects, rotating relatively each other

Let total mass of two bodies is equal to $M$, and distance between them is equal to $2R$. It is possible to write down at once for energy and angular moment:

$$E = \frac{M}{\sqrt{1-\beta^2}}, \qquad (9)$$

$$|\vec{J}| = \frac{R \cdot M \cdot v}{\hbar c^2 \sqrt{1-\beta^2}} \qquad (10)$$

and

$$\frac{R}{c\hbar} \cdot \beta E = \sqrt{J(J+1)} \quad \text{or} \quad \frac{R}{c\hbar} \cdot \sqrt{E^2 - M^2} = \sqrt{J(J+1)} \quad \text{or}$$

$$\frac{R}{c\hbar}\left(\frac{E - M\sqrt{1-\beta^2}}{\beta}\right) = \sqrt{J(J+1)}. \qquad (11)$$

In this case β=υ/c both the angular moment and the energy with the speed increase can grow infinitely.

At small speeds of rotation the dependence (11) in absence of the relativistic increase of mass turns into the known in atomic nucleus physics correlation (1). Really, replacing $(\beta E)^2$ on $E^2 - M^2 = (E + M) \cdot (E - M)$ in expression (11), raised to the second power, and expanding (9) in a series of small β, after its exception we will receive:

$$\frac{2R^2 M}{c^2 \hbar^2}(E - M) = J(J+1),$$

or, in the already ordinary form –

$$E = M + \frac{c^2\hbar^2}{2MR^2}J(J+1).$$

At large speeds of rotation, at β > 0.97, when $E/M > 4$, as calculations show, the linear dependence between energy and moment (see fig. 1) is observed.

## 3. One-particle resonances

### 3.1. Baryon resonances

Among the masses of baryon resonances of a various type it is possible to allocate rotational bands. The characteristic feature of these bands is the substantial increase of masses (more, than twice) and the linear dependence between square of the mass and the orbital moment.

Any of the forms of rotating bodies (without any change of sizes), considered in item 2, does not correspond to observable features. The linear dependence of the energy square on the moment is observed only for a rotating string, but the mass excess for such a body can not be more, than π/2. Others rotating bodies can have more significant mass, excess but there is not the characteristic dependence between the energy and the orbital moment, as for baryon resonances.

It is reasonable to assume, that baryon resonances have the form of a sphere with the uniformly distributed matter density or a string with the squared distribution of mass, depending on the radius. For such bodies the dependence of the energy and the orbital moment on the speed of rotation is identical. At large speeds of rotation, at β > 0.9, there a linear dependence between $E$ and the moment can be. But such a dependence will be in case of a body with constant radius. If in the process of the rotation speed increase the radius will be increased, the dependence between the energy and the angular moment will change (as the orbital moment will be increased) and can reach squared form, as well as it is observed at baryon resonances.

Let us present results of radii calculations of some baryon resonances (see Table 1). The procedure of calculations was the following: β was calculated by (6) for each resonance, then the expression in brackets for the moment (7) was defined. As resonances spins are known, in (7) all values are determined, except for radius, which was calculated by the formula:

$$R = \frac{4}{3} \cdot \frac{\sqrt{J(J+1)}}{(E - M\sqrt{1-\beta^2})} \cdot 198 \cdot \beta.$$

Radii are expressed in $f$, spins - in $\hbar$, and energy - in $MeV$.

As we see, when resonances energy increasing their radii are increased. It is easy to receive extrapolated radii values for masses- founders of rotary bands from Fig.2, where dependences of calculated radii on the moment are shown. So, for protons we have $R = 0.45\ f$ and for isobar (1236) - $R = 0.63\ f$.

## 3.2. Meson resonances

It is possible to allocate rotary states also among meson resonances. So, the sequence of spins and parities of *K*- meson resonances supposes such an interpretation. The calculation of radii by the same equations, which used in a case of baryon resonances, shows their constancy. In contrast to baryons, *K*-meson resonances, despite of the substantial mass growth, do not "inflate" (see Table 2).

## 4. Two-particle resonances

### 4.1. Dibaryon resonances

There are many publications concerning the dibaryon resonances problem and two big reviews /2,3/ among them. The basic part of experimental data of dibaryon resonance masses, which we used, composed mass values from the work /4/. In this work by using of one method of data processing from neutron-proton interactions, where dibaryon resonances should be shown in the purest kind, 17 of proton-proton (in opinion of the authors) resonances is received. Except these data, averaged mass values of 15 resonances from 60 publications in the review /3/ and three resonances from /5/ are involved. This last reference is allocated from others because a belonging of resonances to the rotary band is found out there for the first time. Unfortunately, all experimental data include mass values only. Information about spins is absent.

It has appeared possible to separate all resonances on three classes: NN, NN$\pi$ and $\Delta$N- dibaryons. Each class of dibaryons consists of rotary bands, based on the sum of masses of particles of the appropriate class. This classification is shown in Fig.3, where masses are arranged depending on supposed spins [exactly, on *J (J+1)*]. As resonance masses are not great in compare to the rest mass, it is possible to use (1) for analyses. Supposed spectroscopic information about resonances from /5/ is given by the authors of this work; they have noticed, that these resonances form a part of a rotary band, based on the sum of masses of two nucleons and a pion.

The class of NN - dibaryons has four rotary bands according to four possible quantum states of two nucleons with four various moments of inertia. Values of the mutual distance of two nucleons in a dibaryon resonance, defined from band slopes, are: 2.5; 2.9; 3.2; 3.8 *f* and correspond to peripheral nuclear interactions, where Coulomb and centrifugal potentials dominate.

To each rotary NN-band we have compared a mutual arrangement of the orbital moment and spins of two nucleons, considering well known dependence of nuclear interaction forces on mutual orientation of moments.

In the band, corresponding to the biggest rotating radius, there are resonances, forbidden in the frame of strong interactions for reasons of the symme-try of two-proton system: $^3D_2$, $^3G_4$ and $^3I_6$. The fact of their display can be explained by the small contribution of nuclear forces at such large distance (3.8 $f$).

In the scheme of proton-proton resonances there are vacant places, not defined in presented publications: $^3P_2$, $^3F_2$ и $^1D_2$ with masses 1904, 1910 and 1919 MeV.

If the offered classification is correct, it follows from there, that all resonances with the identical mutual orientation of the orbital moment and spins, irrespective of the complete moment, are formed at the same distance between two protons. There is one radius of rotation for each of four possible orientations.

For trivial explanation of this fact the rectangular form of nuclear potential (to be exact, about sharp edge of potential) can be assumed. For each orientation a depth, and above all, radius of potential wall are characteristic. Then together with electromagnetic and centrifugal potentials we will have resulting potentials shown in Fig.4 and 5, with characteristic jumps. The scattering on such potentials results in resonant situations. These figures are similar to barriers considered by Kudryavtsev A. E. and Obrant G.Z. in /6/. Thus, the first situation (Fig.4) is similar to the scattering on a barrier (elementary amplitude $|A_1|$), and the situation on Fig.5 is similar to scattering on two barriers (total amplitude $|A|$).

This trivial assumption concerning the form of nuclear potential is not compatible with the modern representation about the character of nuclear forces in the theory of singlepion exchange between nucleons.

More plausible reason of spasmodic behaviour of potentials, in our view, is the assumption that the well known fact of spin flip at scattering occurs at the certain distance, characteristic to each orientation. For two nucleons four mutual orientations of the orbital moment and spins are possible, and the reorientation in these states occurs at the certain four various distances.

The reason of spin flip at fixed distance can be the following: at large internucleons distances, where nuclear interaction is weak, the electromagnetic forces set out spins in one of fourth possible positions. The energy of electromagnetic interaction of spins with a magnetic field formed by moving charges, is increased as $R^{-3}$. As nucleons drawing together, nuclear forces become apparent. A part of nuclear potential responsible for reorientation, increases faster, than $R^{-3}$ and at certain distance becomes equal to the energy of electromagnetic interaction. At this distance nuclear spin-orbital interaction reorientates spins.

The energy of nucleon electromagnetic interaction responsible for spin orientation is:

$$U_{elm} = (\vec{\mu}_S \vec{H}_L) = g \frac{e^2}{mc^2} \cdot \frac{\hbar^2}{mR^3} (\vec{l}\vec{s}).$$

Phenomenological nucleon-nucleon potential, including spin-orbital component, has been taken from /7/ (Hamada-Johnson potential):

$$V_{ls} = 2.77 \cdot 3.65 \cdot \left[\frac{exp(-x)}{x}\right]^2 (\vec{l}\vec{s}),$$

where $x = \dfrac{R}{1.43\,\text{f}}$, figures are expressed in *MeV* and are taken for triplet even states.

Behaviour of $U_{\text{elm}}$ and $V_{ls}$ is shown in Fig.6. On absolute values they become equal to each other at internucleon distance $R = 4.54\,f$, that well corresponds to values, determined at the formation of dibaryon resonances systematisation.

However, it is necessary to take into account the following two circumstances. The conservation law of parity forbids transitions $J \to J \pm 1$ and back, and in case of scattering of identical particles there is an additional rule of selection forbidding singlet-triplet transitions. Therefore, effects of dibaryon resonances production are small and are shown in such a degree, in which the laws of strong interaction are violated, or when the contribution of strong interactions is small in comparison with others forms of interactions.

### 4.2. Dipion resonances

The analysis of dipion resonances is inconvenient, since spins are not known for them. Let us try, however, to give the prospective plan of their production and spectroscopy. Mass values of dipion resonances can exceed the sum of masses of two resting π-mesons more, than in five times, therefore, from considered configurations for them one suits only: two rotating bodies separated by certain distance. If the distance between them does not change, linear dependence (11) between the moment and the product $\beta E$ for a rotary band will be observed. As π-mesons have not their own spin, the rotary band should be one (if not consider internal structure of particles).

However, it was not possible to reduce all known dipion resonances (we have taken data from /8/ and /9/) to one rotary band. Most probably, there are four of them (see Fig.7). The linear dependence between the moment and the product βE specifies an invariability of interpion distance. Straight lines slope defines this distance. For four different bands these distances are : 1.94, 2.29, 2.72 and 3.68 *f*.

The presence of several rotary bands testifies to influence of internal structure of π-mesons on formation of resonances. Probably, at such small distances two quarks and two antiquarks are orientated differently for each band and form structures with different moments of inertia.

### 5. Conclusion

Mutual dependences of mass increase and orbital moment for rotating bodies of various configuration have been received and they were used at the analysis of meson and baryon resonances, forming rotary bands.

As a result for one-particle resonances their radii and their possible changes with the spin value increase have been calculated. Thus, in particular, a proton radius equals to 0.45 *f*, and a radius of Δ-isobar – 0.63 *f*.

Dinucleon and dipion resonances are submitted as two rotating bodies. All resonances are placed on some possible rotary bands with various moments of inertia.

Dibaryon resonances are produced as a result of the phenomenon of spin flip (nucleons spins flip relatively to the orbital moment) at nucleon-nucleon peripheral interactions. That fact, that dibaryon resonances is formed rotary bands (four bands according to four possible quantum states), testifies to the correlation of the moments scheme after the spins flip and the distance, at which it is occurred. Each of four possible moment orientations takes place at spin flip at the characteristic distance. The reason of the reorientation of nucleon spins can be the equality at certain distance of a spin-orbital part of nuclear nucleon-nucleon potential and electromagnetic energy of spins interaction with the magnetic field of moving charge. The presence of several rotary bands at dipion resonances can be caused by various polarisation of quarks and antiquarks in the system.

Submitted here spectroscopy of these resonances and plans of their production substantially speculative. For the confirmation of all this picture the data on resonances spins are required, except for their mass values.

# CAPTURES

**Fig.1.** The dependence of energy of states on the orbital moment for bodies of various configurations.

**Fig.2.** Radii of baryon resonances.

**Fig.3.** Three groups of rotating bands of dibaryon resonances: NN, NN$\pi$ and $\Delta$N. Experimental data are taken from /3/, /4/ and /5/. Resonances /4/ are marked by figures in order of mass increase. Mutual orientations of orbital (long arrows) and spin (short arrows) moments are shown for NN bands.

**Fig.4.** The scheme of resonance forming on the repulsing wall of potential. The thing line is the centrifugal potential, the thick one – the resulting potential (centrifugal plus nuclear). At the distance 3.8 fm the reorientation of moments into states, shown on figure, is occurred.

**Fig.5.** The scheme of resonance forming on the potential hole, produced by the nuclear repulsive forces change into the nuclear attractive forces. At the distance 2.5 fm the reorientation of the moment into state, shown on figure, is occurred.

**Fig.6.** The comparison of the energy of electromagnetic interaction to the nuclear (spin-orbital) interaction (1 - electromagnetic interaction, 2 - nuclear spin-orbital interaction).

**Fig.7.** The rotating band of dipion resonances.

**Table 1.** Calculated radii of baryon resonances of a various type.

**Table 2.** Radii of *K*-meson resonances, forming a rotary band.

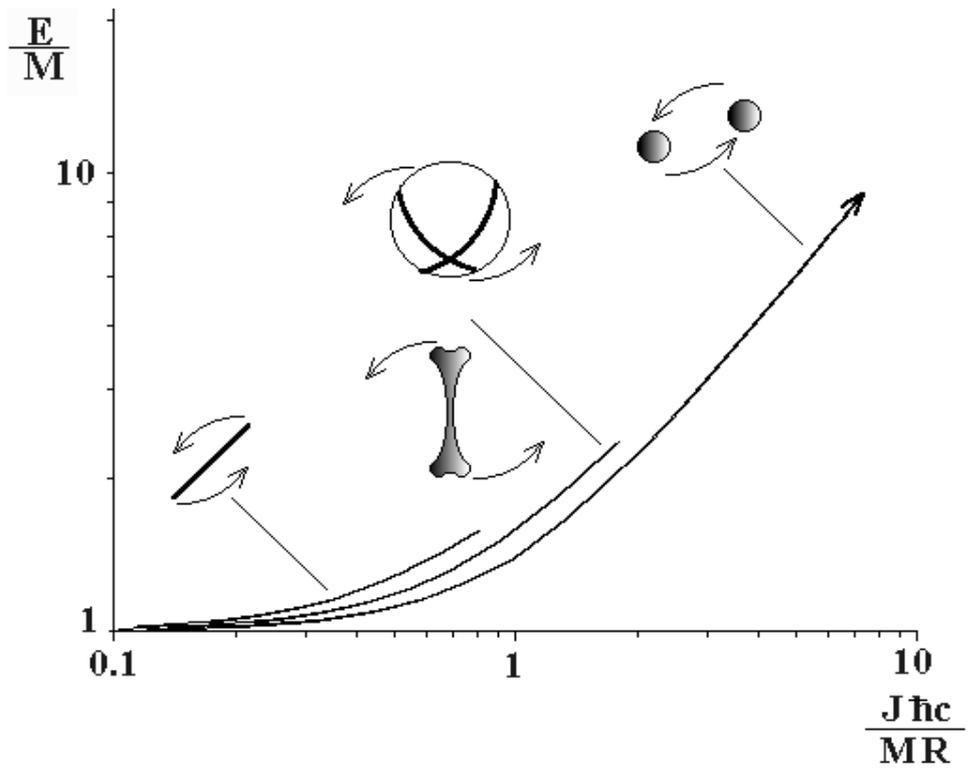

**Fig.1**

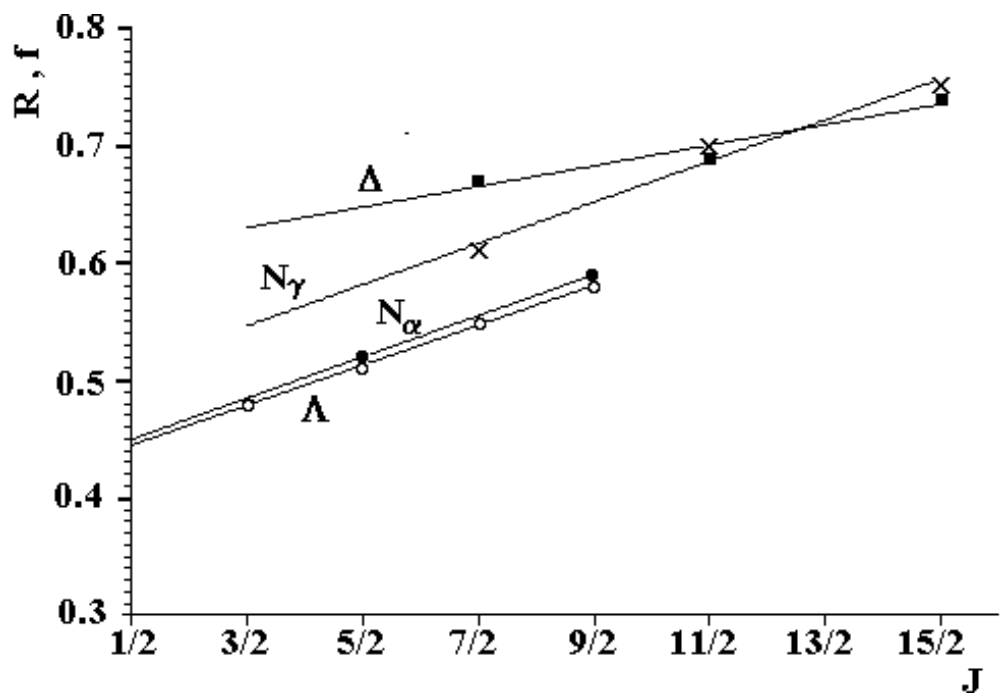

**Fig.2**

Fig.3

Fig.4

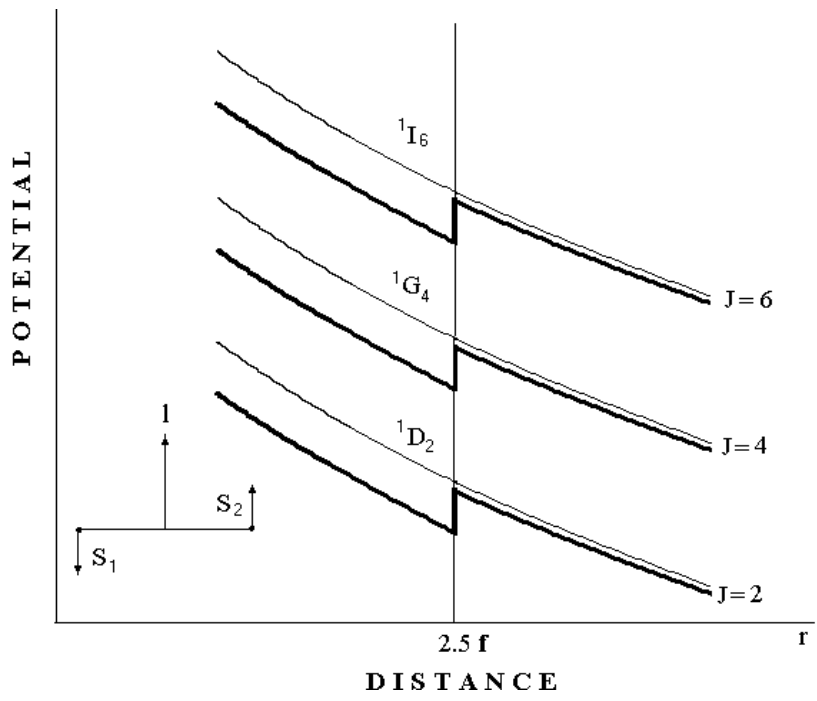

**Fig.5**

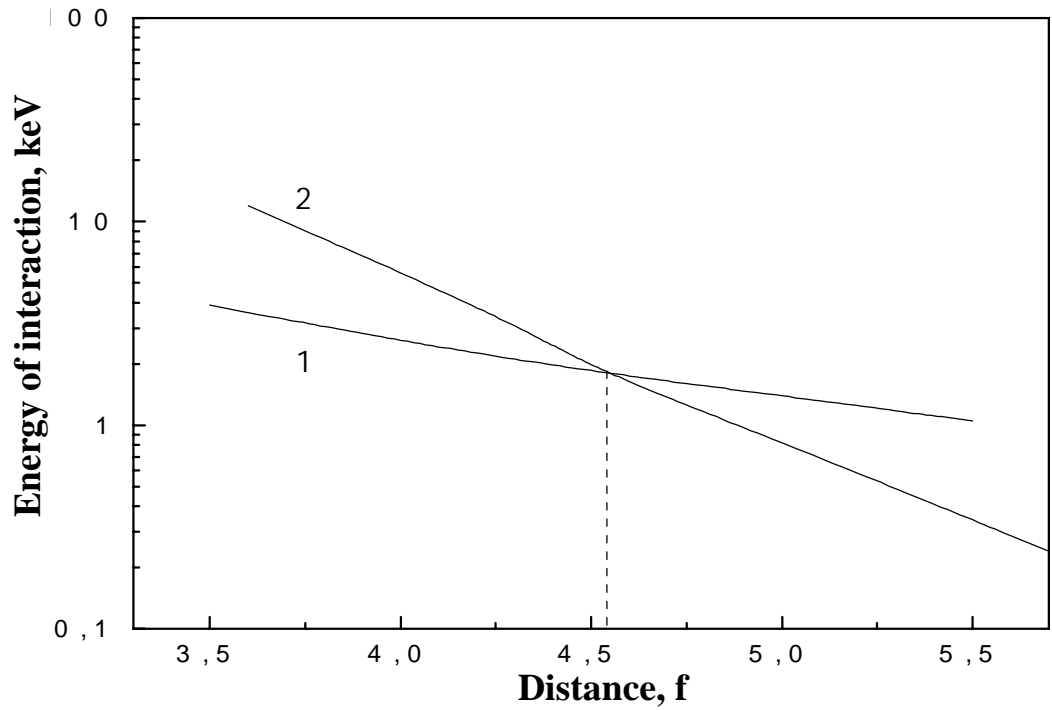

**Fig.6**

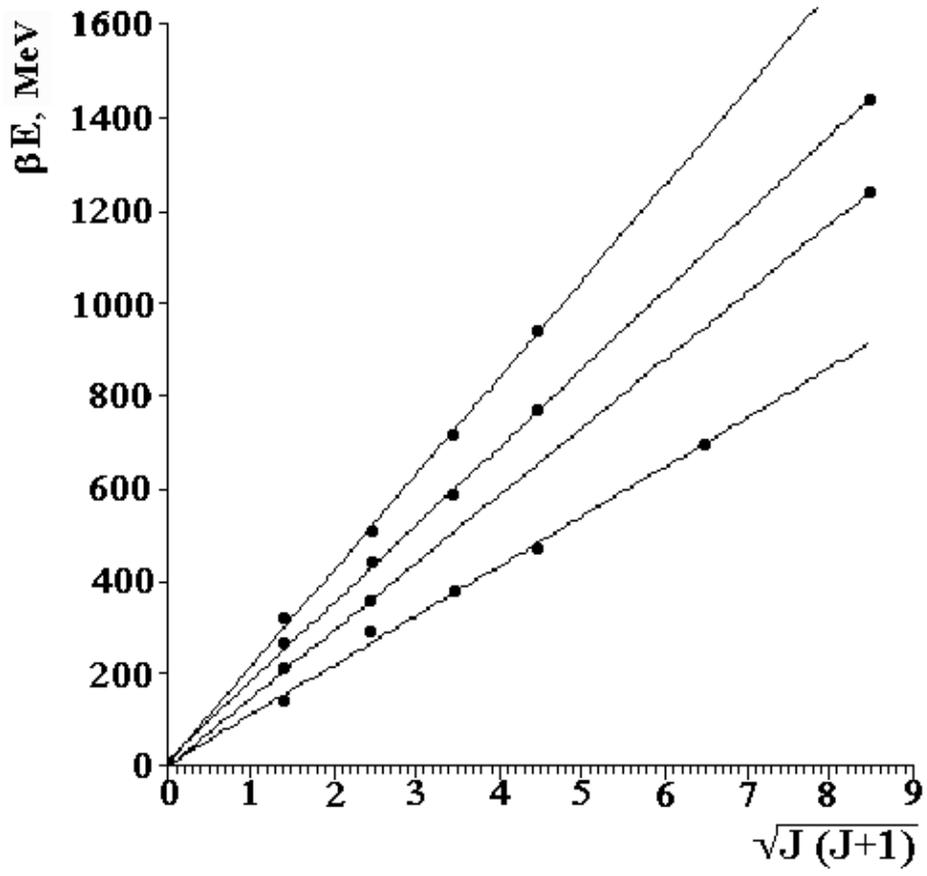

**Fig.7**

**Table 1**

| E, MeV | β | J, ℏ | R, f |
|---|---|---|---|
| Λ-baryons | | | |
| 1115(*M*) | 0 | ½ | |
| 1520 | 0,836 | 3/2 | 0,48 |
| 1815 | 0,940 | 5/2 | 0,51 |
| 2100 | 0,981 | 7/2 | 0,55 |
| 2350 | 0,996 | 9/2 | 0,58 |
| $\Delta_\delta$-baryons | | | |
| 1236(*M*) | 0 | 3/2 | |
| 1924 | 0,920 | 7/2 | 0,67 |
| 2450 | 0,989 | 11/2 | 0,69 |
| 2840 | 0,999 | 15/2 | 0,74 |
| $N_\alpha$-baryons | | | |
| 938(*M*) | 0 | 1/2 | |
| 1688 | 0,970 | 5/2 | 0,52 |
| 2220 | 0,999 | 9/2 | 0,59 |
| $N_\gamma$-baryons | | | |
| 1512(*M*) | 0 | 3/2 | |
| 2210 | 0,886 | 7/2 | 0,61 |
| 2600 | 0,959 | 11/2 | 0,70 |
| 3000 | 0,989 | 15/2 | 0,75 |

**Table 2**

| E, MeV | β | J, ℏ | R, f |
|---|---|---|---|
| 890(M) | 0 | 1 | |
| 1420 | 0,932 | 2 | 0,55 |
| 1780 | 0,990 | 3 | 0,55 |